\documentclass[%
 reprint,
 amsmath,amssymb,
 aps,
]{revtex4-2}

\usepackage{graphicx}

\usepackage{xcolor}

\newcommand{\rf}[1]{(\ref{#1})}

\newcommand{\beq}{\begin{equation}}
\newcommand{\eeq}{\end{equation}}
\newcommand{\beqr}{\begin{eqnarray}}
\newcommand{\eeqr}{\end{eqnarray}}
\newcommand{\lb}[1]{\label{#1}}
\newcommand{\bc}{\begin{center}}
\newcommand{\ec}{\end{center}}
\newcommand{\ct}[1]{\cite{#1}}

\begin{document}

\title{Increased output of superradiant light-emitting diodes due to population fluctuations }


\author{Igor E. Protsenko}
\email{procenkoie@lebedev.ru}

\author{Alexander V. Uskov}

\affiliation{%
P.N.Lebedev Physical Institute of the RAS,
Moscow 119991, Russia
}%

\begin{abstract}
The quantum nonlinear Maxwell-Bloch equations for a single-mode laser with a two-level active medium are solved in the LED regime without adiabatic elimination of the medium polarization, when the population fluctuation spectrum is much narrower than the radiation and polarization   spectra. It is shown that population fluctuations significantly increase the output power and collective Rabi splitting of a superradiant LED. 
\begin{description}
\item[Keywords] Super-radiance, nanolasers, quantum noise
\end{description}

\end{abstract}

\maketitle

\section{\label{Sec1}Introduction}
There is a great demand, interest and progress in the development, study, and application of miniature lasers such as semiconductor lasers \ct{10232115,6261338,Liang2015} and nanolasers \ct{Ma2019,https://doi.org/10.1002/adma.202001996}, in particular, surface-emitting \ct{6365731}, photonic crystal \ct{FITSIOS201897}, microring \ct{Wong2021}, nanofiber \ct{Li2011},  plasmonic \ct{Noginov2009, Noginov2018}  lasers and related devices. 
Theoretical modeling of miniature lasers requires an extension to laser theory, and this paper contributes to this extension. We briefly outline the features of the small lasers we focus on, our previous results and the new contributions made in this paper.  

One important feature  
is a {\em cavity superradiance}, which appears at a large medium-field coupling in a small, low  quality (bad) cavity \ct{Khanin, Belyanin_1998, Koch_2017}.  Many small lasers are superradiant (SR lasers) \ct{Jahnke,PhysRevX.6.011025,PhysRevA.96.013847,PhysRevA.81.033847,PhysRevA.98.063837,Bohnet}.  In small SR lasers, 
the relaxation rates of the field, the polarization, and the field-medium coupling rate are often similar, meaning the dynamics of all variables in the laser equations are important and must be included in the theory. Consequently,  the theory of small SR lasers differs from the standard theory of semiconductor lasers, in which the active medium polarization is eliminated adiabatically \ct{Coldren}. The general motivation behind our theory is to treat the polarization as a dynamical variable.  Polarization dynamics in SR lasers reveal new phenomena such as collective Rabi splitting (CRS) \ct{Andre:19}.
 
Another important feature of small SR lasers is a their quantum radiation. When there are only a few photons in a small cavity, quantum analysis of lasing is required. The theory of a small SR laser must therefore solve quantum nonlinear equations with at least three dynamical variables: the field, the polarization, and the population of states of the active medium. The problem of solving the quantum nonlinear laser equations is non trivial, and much theoretical work  has been done on it in various approximations,  for example Refs. \ct{Andre_opt_expr,  PhysRevA.47.1431,PhysRevA.75.013803,10.1063/1.3697702, doi:10.1063/1.5022958, PhysRevA.111.L011501,LOZING2024116061}. 

Solving quantum nonlinear equations is more challenging than solving classical equations. A detailed comparison of our method with others would require a separate paper.   A brief comparison of our method with other approaches is provided below. Consider the master equation with the Bogoliubov-Born-Green-Kirkwood-Yvon (BBGKY) hierarchy \ct{LOZING2024116061}. In the BBGKY hierarchy, the zero-order approximation neglects field-medium correlations. In our approach, however, such correlations appear already in the zero-order approximation. We believe this is more convenient for calculations. The Heisenberg equations that we use allow for straightforward calculations of observables. They have classical analogues and are more physically transparent than the density matrix equations of Ref.\ct{LOZING2024116061}. We are working
in the frequency domain, so we do not need to integrate in the time domain, which simplifies the calculations. However, in contrast with Ref.\ct{LOZING2024116061}, we cannot describe non-stationary
processes.

Unlike the quantum rate equation (QRE) theory of \ct{PhysRevA.111.L011501} and related papers, we do not need the adiabatic elimination of polarization. However, the QRE is very convenient for numerical simulations, whereas a numerical method based on our approach has yet to be developed.

Our theory is similar to the cluster expansion method (the quantum analog of the cumulant-neglect closure \ct{Wu_1984,10.1115/1.3173083}). In fact, we perform the cluster expansion in the frequency domain. This allows us to describe phenomena, such as CRS, in field spectra. Unlike our approach, the method of Refs. \ct{Jahnke,PhysRevA.75.013803} calculates mean values and correlations, not spectra.

In our papers \ct{PhysRevA.59.1667, Protsenko_2021, Andre:19, PhysRevA.105.053713, https://doi.org/10.1002/andp.202200298, https://doi.org/10.1002/andp.202400121} and the present paper, we use Heisenberg  equations for SR laser (the quantum Maxwell-Bloch equations with Langevin forces).  We make 
the Fourier expansion of the operators, which 
has been used in many papers and books, e.g. Refs.\ct{10.1063/1.105443,RevModPhys.68.127,Scully}. The first paper of our approach is Ref.\ct{PhysRevA.59.1667}, where  a thresholdless laser is described neglecting population fluctuations (PF): it is a zero-order approximation to PF. The zero-order approximation allows one to 
predict collective Rabi splitting, \ct{Andre:19}. 
Following on from Ref. \ct{PhysRevA.59.1667}, our aim is to take the effect of PFs into account in SR laser theory. The PF produces fluctuations in the number of dipoles and in the polarization, and therefore contributes to the radiation.
 
 In Ref. \ct{Protsenko_2021}, within the framework of traditional linear perturbation approach, we analyze the effect of PF above the SR laser threshold  and predict sideband peaks, caused by PF, in the laser field spectrum.   In Ref. \ct{PhysRevA.105.053713} we take into account a weak PF effect below laser threshold, 
in the LED regime, by considering PF as a perturbation. In Ref.   
\ct{https://doi.org/10.1002/andp.202200298} we show that zero-order approximation laser equations are equivalent to the set of equations for normal and inverted harmonic oscillators.   Using results of Ref. \ct{https://doi.org/10.1002/andp.202200298} we  show  
that PF leads to the superthermal photon statistics of a quantum SR LED, as $n\rightarrow 0$  \ct{
https://doi.org/10.1002/andp.202400121}. 

The present work builds on the research of Refs. \ct{PhysRevA.105.053713, 
https://doi.org/10.1002/andp.202400121}
on the PF effect in LEDs. The PF effect on LED power
is not enough suggested, possibly due to the absence of a suitable theoretical model. Here, we extend the work of Refs. \ct{PhysRevA.105.053713, 
https://doi.org/10.1002/andp.202400121} because the perturbative approach of Refs. \ct{PhysRevA.105.053713}  describes only a weak PF effect. In Ref.  \ct{
https://doi.org/10.1002/andp.202400121}, however, we observe a significant increase in SR LED power due to PF when $n$, the number of photons in the LED, is small ($n \rightarrow 0$). This naturally raises the question: is a significant increase in SR LED power due to PF possible for not  small $n$? This paper is motivated by the desire to answer this important question. Here, we will investigate whether a significant increase in SR LED power by PF is possible when $n$ is not small. In such a case, PF is not a perturbation and the approach of Ref. \ct{PhysRevA.105.053713} cannot be used. This motivates us to develop a non perturbative procedure for solving the nonlinear LED equations. This  procedure will be useful for modeling various quantum optical devices. Another motivation was practical: increasing the SR LED power by PF is important for developing efficient miniature light sources. This paper also aims to investigate the effect of PF on collective Rabi splitting in SR LEDs. CRS was considered in Ref. \ct{Andre:19} without PF. Here, we will demonstrate that PF increases CRS.

This paper presents three  findings. First, we develop  a non-perturbative procedure for solving nonlinear quantum Maxwell-Bloch equations with Langevin forces for lasers and quantum optical devices when the population fluctuation spectrum is much narrower than the laser field spectrum. We demonstrate that PFs are a significant source of radiation in the SR LED, increasing its output power by up to 2.5 times. This is the second result. Third, we demonstrate that PF increases the collective Rabi splitting in the LED spectrum. 

Section \ref{Sec2} describes the LED model. The procedure for solving the equations of Sec. \ref{Sec2} is described in Sec. \ref{Sec2a}, using results from Refs. \ct{PhysRevA.105.053713, https://doi.org/10.1002/andp.202400121}. Here we take into account the dependence of PF on the cavity photon number $n$, which was neglected in Ref. \ct{ https://doi.org/10.1002/andp.202400121} as $n\rightarrow 0$ and considered as a perturbation in Ref. \ct{PhysRevA.105.053713}. The effect of the PF on the output power and the field spectra of the LED will be demonstrated in Sec.  \ref{Sec4}. There we compare the effect of PF on non-SR and SR LEDs, formulate conditions for the maximum increase in LED output power and CRS due to PF, and discuss the results. The paper ends with a conclusion. 
\section{\label{Sec2} LED model}
\begin{figure}[thb]\bc
\centering
\includegraphics[width=7.5cm]{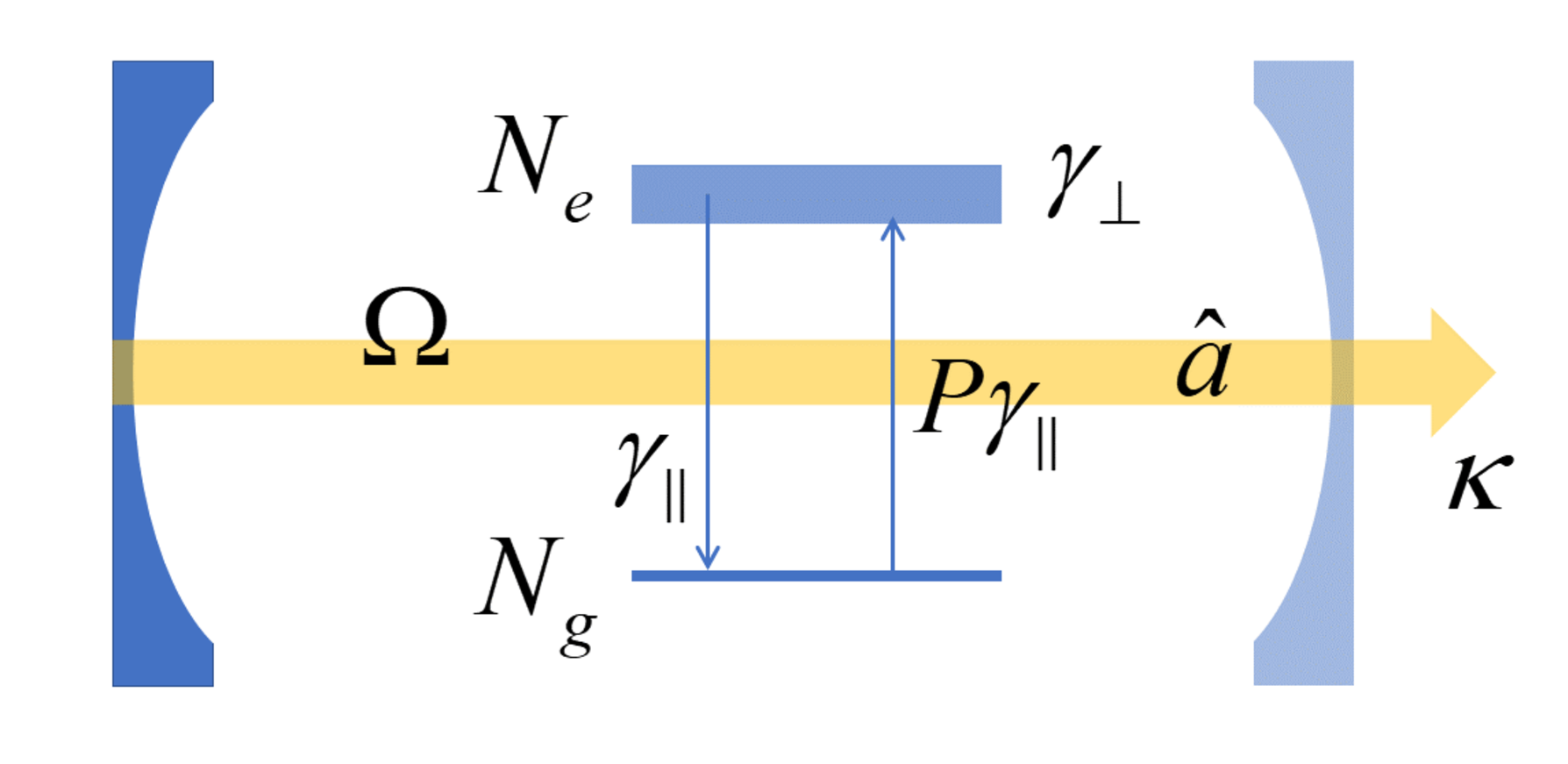}
\caption{The scheme of two-level LED model. The field with Bose operator $\hat{a}$ interacts with $N_0$  two-level  atoms (quantum dots) with mean the  upper (lower) level  populations $N_e$ ($N_g$), the transition width $\gamma_{\perp}$ and the medium-field coupling constant $\Omega$. The upper state decay rate is $\gamma_{\parallel}$ and the pump rate is $P\gamma_{\parallel}$.  The cavity field decay rate is $\kappa$.       }
\label{Fig0}\ec
\end{figure}

We consider a stationary regime of the single-mode LED, shown in Fig.~\ref{Fig0},  with the active medium of $N_0$  two-level emitters.   
%
%
%

The cavity mode and the transitions of the emitters are in resonance.  The LED is described by the Heisenberg-Langevin equations, derived in Ref. \ct{PhysRevA.105.053713}
\begin{subequations}\lb{MBE_0}\beqr
  \dot{\hat{a}}&=&-\kappa \hat{a}+\Omega \hat{v}+\sqrt{2\kappa}\hat{a}_{in} \lb{MBE_1}\\
 \dot{\hat{v}}&=&-\left( {{\gamma }_{\bot }}/2 \right)\hat{v}+\Omega f\left(\hat{a}N+2\hat{a}{{\delta{\hat{N}}}_{e}} \right)+{{{\hat{F}}}_{v}} \lb{MBE_2}\\
 \delta{{{\dot{\hat{N}}}}_{e}}&=&-\Omega \delta\hat{\Sigma}-\gamma_{\parallel}(P+1)\delta\hat{N}_e +\hat{F}_{N_e}. \lb{MBE_3} 
\eeqr\end{subequations}
Here $\hat{a}{{e}^{-i{{\omega }_{0}}t}}$ is the cavity field operator with the field amplitude Bose operator $\hat{a}$ and the  optical carrier frequency ${{\omega }_{0}}$;  $\hat{v}$ is the active medium polarization amplitude operator $\hat{v}=\sum_{i=1}^{N_0}f_i\hat{\sigma}_i$; $\hat{\sigma}_i$ is the lowing operator of i-th emitter (see Eqs.(2) of Ref. \ct{PhysRevA.105.053713}), $f_i$ describes the difference in the field coupling rates for different emitters;
$\delta {{\hat{N}}_{e}}$ is the medium upper state population fluctuation operator; $\delta \hat{\Sigma }=\hat{\Sigma }-\Sigma $  describes fluctuations of the operator $\hat{\Sigma }={{\hat{a}}^{+}}\hat{v}+{{\hat{v}}^{+}}\hat{a}$ of the dipole interaction between the LED field and the active medium, the mean value of $\hat{\Sigma}$ is $\Sigma $; $\hat{a}_{in}$ is the input vacuum field operator; $\sqrt{2\kappa}\hat{a}_{in}$, ${{\hat{F}}_{v}}$ and ${{\hat{F}}_{{{N}_{e}}}}$ are Langevin force operators;  $\kappa$, ${{\gamma }_{\bot }}/2$ and ${{\gamma }_{\parallel }}$ are the cavity field, polarization, and upper-state population relaxation rates, respectively; $\Omega$ is the Rabi frequency of the field-medium coupling; $P$ is the normalized excitation rate (the pump) of the active medium,   the factor $f=N_0^{-1}\sum_{i=1}^{N_0}f_i^2\approx 1/2$ is due to the averaging over the cavity field and emitter coupling rates, $N={{N}_{e}}-{{N}_{g}}$ is the mean population inversion, ${{N}_{g}}$, (${{N}_{e}}$) are the mean lower (upper) state populations  of the emitters, ${{N}_{e}}+{{N}_{g}}={{N}_{0}}$. We assume $N_0\gg 1$.  

${{N}_{e,g}}$ can be found from the energy conservation law, which follows from the stationary LED equations for the mean values \ct{PhysRevA.59.1667,  PhysRevA.105.053713,Andre:19} 
\beq
2\kappa n={{\gamma }_{\parallel }}\left( P{{N}_{g}}-{{N}_{e}} \right), \lb{En_c_l}
\eeq
where $n=\left\langle {{{\hat{a}}}^{+}}\hat{a} \right\rangle$ is the mean number of photons in the LED cavity, $n =(2\pi)^{-1}\int\limits_{-\infty }^{\infty }{n(\omega )d\omega }$,  $n(\omega )$ is the cavity field spectrum,
\beq
\left\langle {{{\hat{a}}}^{+}}(-\omega ')\hat{a}(\omega ) \right\rangle =n(\omega )\delta \left( \omega +\omega ' \right), \lb{F_sp} 
\eeq
$\hat{a}(\omega )$ is the Fourier-component of $\hat{a}(t)$: 
\beq \hat{a}(t)=(2\pi)^{-1/2}\int\limits_{-\infty }^{\infty }{\hat{a}(\omega ){{e}^{-i\omega t}}d\omega }.\lb{a_Fourier}\eeq 

From Eqs.~\rf{MBE_0}, we find $\hat{a}(\omega )$ as a function of $N_e$;  calculate $n(\omega)$ and $n(N_e)$ using the correlation properties of the Langevin force operators in Eqs.~\rf{MBE_0}, and find $N_e$ by solving Eq.~\rf{En_c_l}. The same procedure for the calculation of $N_e$ has been performed in Refs. \ct{PhysRevA.59.1667,Andre:19} without considering population fluctuations. 

To start with, we look again at the estimates that were used before to solve Eqs.~\rf{MBE_0} in our method, and clarify why a fresh approach is required.
Our aim is to describe the effect of PF $\delta\hat{N}_e$ on LEDs. In Eq.~\rf{MBE_2}, we observe that the nonlinear term, $\sim \hat{a}\delta\hat{N}_e$, contributes to polarization and, consequently, to LED radiation. In the lasing regime, when the radiation line is narrow, $\hat{a}$ can be approximated by a constant, $\hat{a} \approx A$, so $\hat{a}\delta\hat{N}_e\approx A\delta\hat{N}_e$. Then, Eqs. \rf{MBE_0} become linear in the operators and can be solved using the standard perturbation procedure, as described in Ref. \ct{Protsenko_2021}. However, in the present paper, we do not consider the lasing regime, the radiation line is not narrow; thus, the approximation $\hat{a} \approx A$ cannot be used.

In Refs. \ct{PhysRevA.59.1667,Andre:19}  we use a zero-order approximation, which neglects  $\delta\hat{N}_e$. Then, we solve the linearized equations  \rf{MBE_1}  \rf{MBE_2} and investigate thresholdless lasing in Ref. \ct{PhysRevA.59.1667} and collective Rabi splitting for an SR LED in Ref. \ct{Andre:19}.  The non linearity $\hat{a}\delta\hat{N}_e$ complicates the calculations in the LED regime, where the usual linearization, as in Ref.  \ct{Protsenko_2021},  is  invalid. We then apply the first-order perturbation approach in Ref. \ct{PhysRevA.105.053713},  replacing $\hat{a}$ in $\hat{a}\delta\hat{N}_e$ with the zero-order approximation result  from Refs. \ct{PhysRevA.59.1667,Andre:19} and neglecting  $\delta\Sigma$  in Eq.~\rf{MBE_3}  for $\delta\hat{N}_e$.

The perturbation approach requires that the first-order contribution to be smaller than the zero-order contribution. However, this is not always the case for small SR LEDs, as demonstrated in Ref. \ct{https://doi.org/10.1002/andp.202400121}. In Ref. \ct{https://doi.org/10.1002/andp.202400121}, we found that population fluctuations significantly increase SR LED power in the strongly quantum regime, when $n\rightarrow 0$.  From a practical standpoint, it is interesting to analyze whether a strong increase
in LED power is possible when $n$ is not small. In order to perform such an analysis  we introduce a {\em non perturbative} approach, presented in the next section, which takes into account the nonlinear term in Eq.~\rf{MBE_2} in the LED regime. We assume that the LED radiation spectrum is broad, the spectral width of the radiation  is of the order of $\max \left\{ \kappa ,{{\gamma }_{\bot }} \right\}$   and $\kappa ,{{\gamma }_{\bot }}\gg {{\gamma }_{\parallel }}P$ which occurs at typical parameters in the LED regime.  

Using the solution of Eqs.~\rf{MBE_0}, we show that PF significantly increase the LED output power not only for $n\rightarrow 0$, but also when $n$ is not small; PF maintain, and even enhance the collective Rabi splitting \ct{Andre:19} in the LED spectra. 

\section{\label{Sec2a}  calculation procedure}

Equations~\rf{MBE_0} lead to equations for the operators $\hat{a}(\omega)$ and ${\hat{v}}(\omega)$ of the Fourier components of $\hat{a}(t)$ and ${\hat{v}}(t)$
\begin{subequations}\lb{MBE_FC}\beqr
 (\kappa-i\omega)\hat{a}(\omega)&=&\Omega \hat{v}(\omega)+\sqrt{2\kappa}\hat{a}_{in}(\omega) \lb{MBE_FC_1}\\
  ({{\gamma }_{\bot }}/2-i\omega){\hat{v}}(\omega)&=&\\& &\nonumber\hspace{-1cm}\Omega f\left[\hat{a}(\omega)N+2(\hat{a}{{\delta{\hat{N}}}_{e}})_{\omega} \right]+{{{\hat{F}}}_{v}}(\omega) \lb{MBE_FC_2}
\eeqr\end{subequations}
where ${{\left( \hat{a}\delta {{{\hat{N}}}_{e}} \right)}_{\omega }}$ is the Fourier component of  $\hat{a}\delta {{{\hat{N}}}_{e}}$. 
From Eqs.~\rf{MBE_FC_1} and \rf{MBE_FC_2} we find 
\beq\hat{a}(\omega )=\frac{2{{\Omega }^{2}}f{{\left( \hat{a}\delta {{{\hat{N}}}_{e}} \right)}_{\omega }}+\Omega {{{\hat{F}}}_{v}}(\omega )+\left( {{\gamma }_{\bot }}/2-i\omega  \right)\sqrt{2\kappa}\hat{a}_{in}(\omega )}{s(\omega)},\lb{a_omega}\eeq
where $s(\omega)=\left( \kappa -i\omega  \right)\left( {{\gamma }_{\bot }}/2-i\omega  \right)-\kappa {{\gamma }_{\bot }}N/2{{N}_{th}}$ and ${{N}_{th}}=\kappa\gamma_{\perp}/2\Omega^2f$ is the threshold population inversion found in the semiclassical laser theory \ct{SLW}. We substitute $\hat{a}(\omega )$ from Eq.~\rf{a_omega} into Eq.~\rf{F_sp} and calculate the field spectrum
\beq
n(\omega )=\left[{{{\left( 2{{\Omega }^{2}}f \right)}^{2}} S_{a{{N}_{e}}}(\omega )+{{\gamma }_{\bot }}f{{\Omega }^{2}}{{N}_{e}}}\right]/{|s(\omega )|^2},\lb{lf_spectrum}
\eeq
where $S_{a{{N}_{e}}}(\omega )$ is a convolution
\beq
S_{a{{N}_{e}}}(\omega )=\frac{1}{2\pi }\int\limits_{-\infty }^{\infty }{\left[ n(\omega -\omega ')+c(\omega -\omega ')/2 \right]{{\delta }^{2}}{{N}_{e}}(\omega ')d\omega '},\lb{S_aNe}
\eeq
\beq
c(\omega )=\left[2\kappa {{\omega }^{2}}+(\kappa \gamma _{\bot }^{2}/2)(1-N/{{N}_{th}})\right]/{|s(\omega )|^2}\lb{c_omega}
\eeq
and ${{\delta }^{2}}{{N}_{e}}(\omega )$ is the power spectrum of PF.  For the derivation of Eqs.~\rf{lf_spectrum}, \rf{S_aNe} we use the power spectrum $2D_{v^+v}(\omega)$ of the $\hat{F}_v$ Langevin force, $\left<\hat{F}_{v^+}(\omega)\hat{F}_v(\omega')\right> = 2D_{v^+v}(\omega)\delta(\omega+\omega')$, found in Refs. \ct{PhysRevA.105.053713, https://doi.org/10.1002/andp.202400121}
\beq
2{{D}_{{{v}^{+}}v}}(\omega )=f{{\gamma }_{\bot }}{{N}_{e}}+2{{f}^{2}}{{\Omega }^{2}}{{\left( c*{{\delta }^{2}}{{N}_{e}} \right)}_{\omega }},\lb{power_sp_v}
\eeq
where ${{\left( c*{{\delta }^{2}}{{N}_{e}} \right)}_{\omega }}$ is a convolution of $c(\omega)$ and ${{\delta }^{2}}{{N}_{e}}(\omega )$. In Eq.~\rf{MBE_FC_1}, $\hat{a}_{in}$ is the vacuum field operator, $\left<\hat{a}_{in}^+(\omega)\hat{a}_{in}(\omega')\right>=0$, so $\sqrt{2\kappa}\hat{a}_{in}(\omega)$ does not contribute to $n(\omega)$. In deriving  \rf{lf_spectrum}, we neglect the correlations between the polarization and population fluctuations, which is an acceptable when $N_0$ is large.  In fact, some of the two photons are most likely emitted by two different emitters with uncorrelated population and polarization fluctuations, as discussed in more detail in   Ref. \ct{https://doi.org/10.1002/andp.202200298}.   

Let us now discuss the power spectrum \rf{power_sp_v} of the polarization fluctuations, which describes {\em the sources} of the radiation. The first term on the right-hand side of Eq.~\rf{power_sp_v} comes from the zero-order approximation \ct{PhysRevA.59.1667,Andre:19}. It describes the polarization fluctuations of excited emitters. The second term in Eq.~\rf{power_sp_v} appears to satisfy $[\hat{a},\hat{a}^+]=1$ when PF is taken into account \ct{PhysRevA.105.053713,https://doi.org/10.1002/andp.202400121}. The second term is physically equivalent to the PF contribution to spontaneous emission in the LED cavity mode. Meanwhile, the nonlinear term $\hat{a}\delta\hat{N}_e$ in Eq.~\rf{MBE_2}  describes the PF contribution to stimulated emission. 

So, when the number $n$ of photons in the LED approaches zero and the stimulated emission is negligible, the PF effect is mainly described by the second term in Eq.~\rf{power_sp_v}, while the nonlinear term $\hat{a}\delta\hat{N}_e$ in Eq.~\rf{MBE_2} is neglected 
\ct{https://doi.org/10.1002/andp.202400121}.  
According to the derivation in Refs. \ct{PhysRevA.105.053713,PhysRevA.105.053713,https://doi.org/10.1002/andp.202400121}, the second term in Eq.~\rf{power_sp_v} is not a perturbation. It may exceed the first term, when $\gamma_{\perp}$ is small and  $\Omega$ is large, as in SR LEDs.
See examples in Ref. \ct{https://doi.org/10.1002/andp.202400121}, when the second term in Eq.~\rf{power_sp_v}  is larger than the first term. The significant contribution of PF at $n\rightarrow 0$ found in Ref. \ct{https://doi.org/10.1002/andp.202400121} prompted the present study of the PF effect, when $n$ is not small.  

We assume that the width of the population fluctuation spectrum $\delta^2{{N}_{e}}(\omega)$ is much smaller than the widths of the $n(\omega)$ and $c(\omega)$ spectra. In the LED regime, this is true if $\gamma_{\parallel} \ll \kappa, \gamma_{\perp}$, which is usually satisfied for semiconductor lasers with fast polarization dephasing and a low quality cavities. Thus,  in Eq.~\rf{S_aNe} we approximate
\beq
S_{a{{N}_{e}}}(\omega )\approx[n(\omega)+c(\omega)/2]{{\delta }^{2}}{{N}_{e}},\lb{S_aNe_nbs}
\eeq
where $\delta^2N_e =(2\pi)^{-1}\int\limits_{-\infty }^{\infty }\delta^2{{N}_{e}}(\omega )d\omega$ is the PF dispersion. Substituting $S_{a{{N}_{e}}}(\omega )$ from Eq.~\rf{S_aNe_nbs} into Eq.~\rf{lf_spectrum} we find explicitly the field spectrum in the LED cavity  
\beq
n(\omega )=f{{\Omega }^{2}}\frac{2{{\Omega }^{2}}{{f}}c(\omega ){{\delta }^{2}}{{N}_{e}}+{{\gamma }_{\bot }}{{N}_{e}}}{|s(\omega )|^2-{{4{{\Omega }^{4}}f^2}}{{\delta }^{2}}{{N}_{e}}},\lb{n_sp_appr}
\eeq
The main difference between the result \rf{n_sp_appr} and the other results  \ct{PhysRevA.105.053713,https://doi.org/10.1002/andp.202400121} is the {\em nonlinear} dependence of $n(\omega)$ on $\delta^2{N}_e$ in \rf{n_sp_appr}. The result of Ref. \ct{https://doi.org/10.1002/andp.202400121} follows from \rf{n_sp_appr} by neglecting the term $\sim\delta^2{N}_e$ in the denominator of \rf{n_sp_appr}. The result of Ref. \ct{PhysRevA.105.053713} can be obtained from \rf{n_sp_appr} using a Tailor expansion of \rf{n_sp_appr} in $\delta^2{N}_e$ up to the  first-order terms. Thus the results of Refs. \ct{PhysRevA.105.053713,https://doi.org/10.1002/andp.202400121} are {\em linear} in $\delta^2{N}_e$. The nonlinearity of \rf{n_sp_appr} in $\delta^2{N}_e$ may lead to new resonances at high pump power ($P> 1$), when $\delta^2{N}_e$  significantly depends on $n$. We will leave the study of Eq.~\rf{n_sp_appr} with a strong pump for the future. Here we will demonstrate that the nonlinearity of Eq.~\rf{n_sp_appr}   significantly amplifies the impact of PF on the field spectra of SR LEDs when  $P$ is about 1. 

The population fluctuation dispersion $\delta^2N_e$ is unknown in Eq.~\rf{n_sp_appr}. We find $\delta^2N_e$  using the procedure described in Refs. \ct{PhysRevA.105.053713,https://doi.org/10.1002/andp.202200298}. There we derive equations for binary operators, such as $\hat{n}=\hat{a}^+\hat{a}$ and $\hat{\Sigma}$, linearize  them respectively to fluctuations and solve them. The procedure of Refs. \ct{PhysRevA.105.053713,https://doi.org/10.1002/andp.202200298} is valid, if a part $(\delta^2N_e)_{f}$ of $\delta^2N_e$ caused by the cavity field is small $(\delta^2N_e)_{f}^{1/2}\ll N_e$, which is satisfied  in this paper. 

In Refs. \ct{PhysRevA.105.053713,https://doi.org/10.1002/andp.202200298},  the simplified expression for $\delta^2N_e$   was used as the mean number of cavity photons $n\rightarrow 0$ and $\delta^2N_e$ does not depend  on $n$.  In contrast to Refs. \ct{PhysRevA.105.053713,https://doi.org/10.1002/andp.202200298}, here $n$ is not small, so that we take into account the dependence of $\delta^2N_e$  on $n$.

The result for $\delta^2N_e$ is a function of $N_e$, which is  an unknown parameter. To find $N_e$, we insert $\delta^2N_e(N_e)$ into Eq.~\rf{n_sp_appr}, calculate the mean photon number $n(N_e)=(2\pi)^{-1}\int_{-\infty}^{\infty}n(\omega)d\omega$, and insert $n(N_e)$ into the law of energy conservation \rf{En_c_l}. We then find $N_e$ by  solving Eq.~\rf{En_c_l}. Substituting $N_e$ into the expression \rf{n_sp_appr} we find the field spectrum $n(\omega)$; integrating $n(\omega)$ over the frequency, we find the mean photon number $n$.

\section{\label{Sec4} Population fluctuation effect on the LED power and spectra}

\subsection{\label{Sec3} LED parameters}

In the examples, we use parameter values close to those typical  for semiconductor microlasers and LEDs with quantum dot active media and photonic crystal cavities.  The wavelength of the cavity field  in vacuum is ${{\lambda }_{0}}=1.55$ $\mu$m; the linear refractive index of the medium is ${{n}_{r}}=3.3$. We take the cavity mode volume ${{V}_{c}}={{n}_{c}}{{V}_{\min }}$, where ${{V}_{\min }}={{\left( {{\lambda }_{0}}/2{{n}_{r}} \right)}^{3}}$ is the minimum volume of the resonant optical cavity and ${{n}_{c}}\ge 1$ is the normalized cavity volume expressed in $V_{\min }$ units.  We have chosen ${{n}_{c}}$ in the region between $2$ and $100$. The dipole momentum of the two-level emitter (quantum dot) optical transition is $d={{10}^{-28}}$ Cm. The Rabi frequency $\displaystyle\Omega =\frac{d}{{{n}_{r}}}\sqrt{\frac{{{\omega }_{0}}}{{{\varepsilon }_{0}}\hbar {{V}_{c}}}}$, $\Omega$ is in the region from $3\cdot 10^{10}$ to $22\cdot 10^{10}$  rad/s depending on $V_c$. We consider two values of the polarization decay rate. The first value is ${{\gamma }_{\bot }}={{10}^{12}}$ rad/s, which is typical at room temperature \ct{PhysRevB.46.15574}. 
 The second value is ${{\gamma }_{\bot }}=5\cdot {{10}^{10}}$ rad/s. 
 This low value of ${{\gamma }_{\bot }}$ can be achieved by cooling the LED, for example. The decay rate of the cavity field is  $\kappa =2.5\cdot {{10}^{10}}$ or  $\kappa =5\cdot {{10}^{11}}$ rad/s, for cavity quality factors of $2.4\cdot {{10}^{4}}$ and $1.2\cdot {{10}^{3}}$, respectively. We consider ${{N}_{0}}=100$ or ${{N}_{0}}=200$ active  emitters in the cavity. The decay rate of the upper-state population of the active medium  is ${{\gamma }_{\parallel }}={{10}^{9}}$ rad/s, which is on the order of the spontaneous emission decay rate for the  optical dipole transition. 

\subsection{\label{Sec3a}  LED power and the field spectra}

We first consider examples with the LED of a small adiabatic parameter $2\kappa /{{\gamma }_{\bot }}\ll 1$, the superradiance gives a small contribution to the radiation of such a non superradiant (non-SR) LED  \ct{Andre:19,PhysRevA.105.053713}. In principle, a non-SR LED can be described by the quantum rate equations \ct{Coldren}, where the polarization is eliminated adiabatically. 

In the first example, we take $\kappa =2.5\cdot {{10}^{10}}$  rad/s, ${{\gamma }_{\bot }}={{10}^{12}}$ rad/s, so $2\kappa /{{\gamma }_{\bot }}=0.05\ll 1$. We consider the values of normalized cavity volume ${{n}_{c}}=100$, 50, 10, 5, and  2, so $\Omega =\left( 0.3,\ 0.43,\ 0.97,\ 1.37,\ 2.17 \right)\cdot {{10}^{11}}$ rad/s, respectively, and  $\Omega \ll {{\gamma }_{\bot }}$ in all cases, but with ${{n}_{c}}=2$, when $\Omega =2.17\cdot {{10}^{11}}\sim {{\gamma }_{\bot }}={{10}^{12}}$ rad/s. Using the differential gain (the spontaneous emission rate to the cavity mode) $g\equiv 4{{\Omega }^{2}}f/(2{{\kappa }_{0}}+{{\gamma }_{\bot }})$, we estimate the factor $\beta \equiv g/(g+{{\gamma }_{\parallel }})\sim 1$ is in the region between $0.64$ and $0.989$. So the LEDs in all examples have significant spontaneous emission to the cavity mode: the LEDs are thresholdless. We consider ${{N}_{0}}=100$ or 200 active emitters in the cavity. We compare results found with the help of  the non-perturbative solution Eq.~\rf{n_sp_appr} with results obtained in approximations of Refs. \ct{PhysRevA.105.053713} and \ct{https://doi.org/10.1002/andp.202400121}.
%
%
\begin{figure}[thb]\bc
\centering
\includegraphics[width=8.5cm]{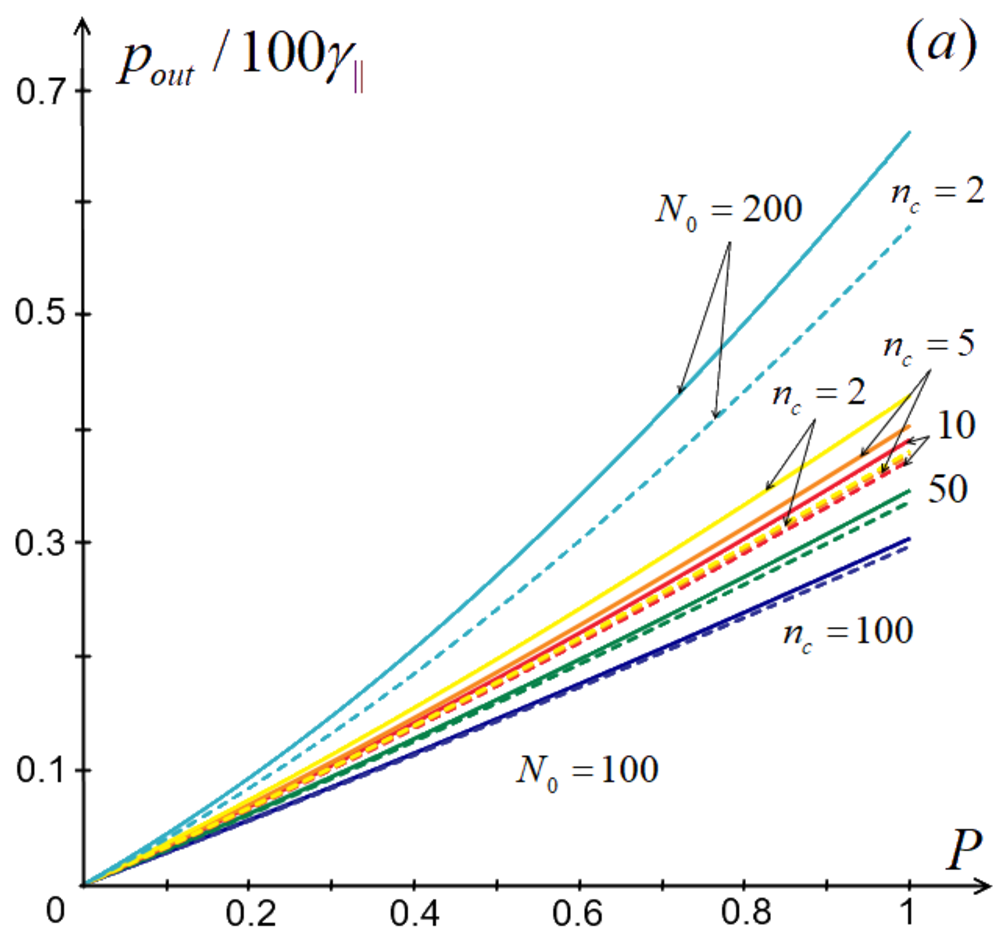}
\includegraphics[width=8.5cm]{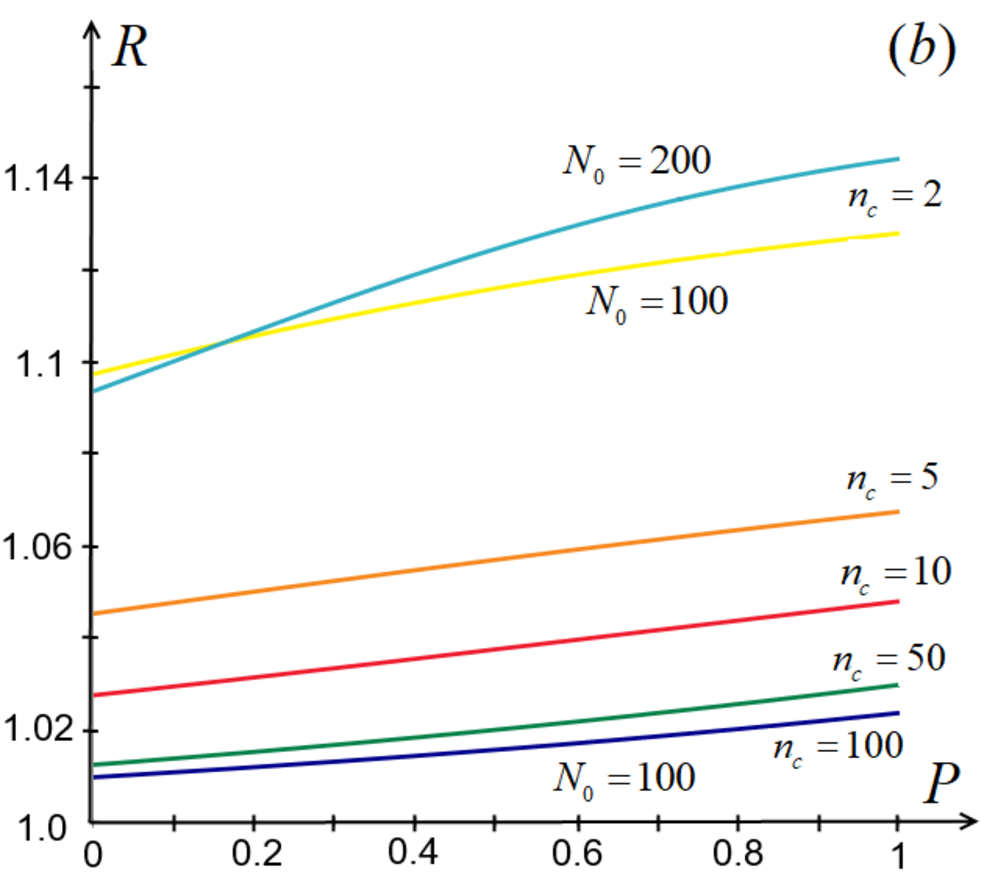}
\caption{(a) The mean output power ${{p}_{out}}$ (in photons/sec) versus the normalized pump $P$ calculated with (solid curves) and without (dashed curves) PF for the non-SR LED with  with $n_c$ values shown near curves. The number of emitters $N_0$=100 for all curves but the highest one with $N_0=200$. (b) The factor $R$ of the increase the LED output power by PF; notations of curves are the same as in Fig.1a. The PF relative contribution $R-1$ is small for all curves.  $R$ grows with $\Omega$, (i.e. for smaller $n_c$), remains about the same with the change of  $N_0$  and gradually increases with $P$, which is explained in the discussion section.     }
\label{Fig1}\ec
\end{figure}
%

Figure \ref{Fig1}(a) shows the mean output power $p_{out}=2\kappa n$ (in photons/sec) calculated with (solid lines) and without (dashed lines) PF for the non-SR LED. We see  a small contribution of PF to the non-SR LED radiation increasing as the cavity becomes smaller (i.e., for smaller ${{n}_{c}}$). Fig.\ref{Fig1}(b)   shows the relative LED output power increase $R={{p}_{out}}/{{p}_{out}}(\delta {{N}_{e}}=0)$ due to PF, where ${{p}_{out}}$ [${{p}_{out}}(\delta {{N}_{e}}=0)$] are the LED powers found with (without) PF. We see that $R>1$, so PF increases the LED power in all cases. The relative contribution $R-1$ of PF to the radiation is small, about a few percent, when the polarization decay rate ${{\gamma }_{\bot }}\gg \Omega $, as for the four lowest curves in Fig.\ref{Fig1}(b). The PF contribution increases with the larger active medium and field coupling (smaller cavity volume and larger $\Omega$), as $\Omega $ approaches ${{\gamma }_{\bot }}$, then $R-1$ reaches $\sim 14\%$ (see the two highest curves in Fig.\ref{Fig1}~b). $R$ gradually decreases with the  pump rate $P$.  Increasing in the number of emitters $N_0$ does not significantly change  $R$. 
%
%
\begin{figure}[thb]\bc
\centering
\includegraphics[width=9.5cm]{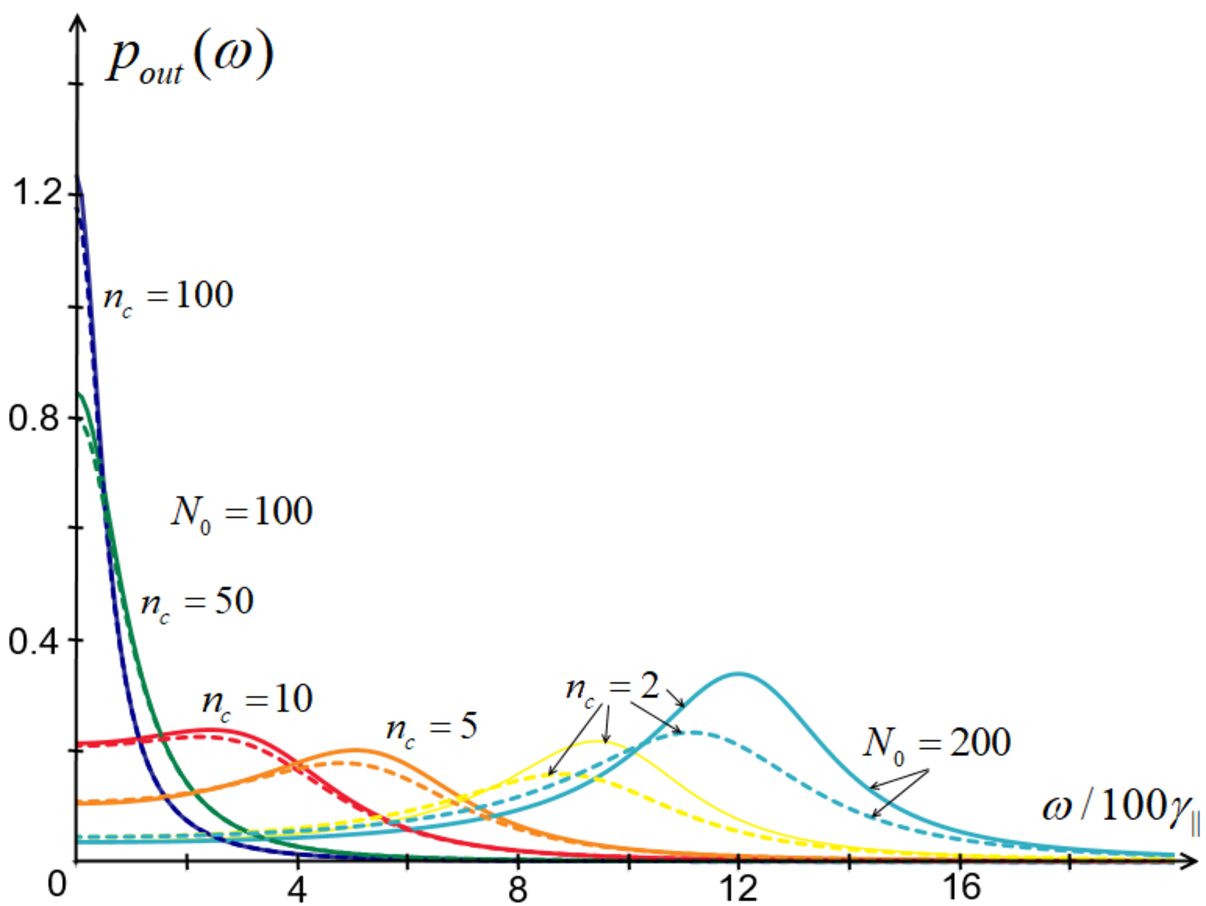}
%
\caption{The output field  spectra for non-SR LED for $P=1$. Parameters for curves are the same as in Fig.\ref{Fig1}.  Solid (dashed) curves are calculated with (without) PF. Values of $n_c$ are shown near curves. Population fluctuations increase and shift the maxima in spectra with CRS to the right  i.e. the CRS increases respectively to the case without PF, as much as smaller is the normalized cavity volume $n_c$. The population fluctuation effect is more visible for larger $N_0$ (compare curves with $n_c=2$ and $N_0=100,200$).}
\label{Fig2}\ec
\end{figure}
%

The output power spectra $p_{out}(\omega) = 2\kappa n(\omega )$ of the non-SR LED are shown in Fig.~\ref{Fig2} for the same parameters as in Fig.~\ref{Fig1}.
Here the PF slightly increases the CRS, e.g. the CRS maxima in curves grow and move to the right much as the normalized cavity volume $n_c$ decreases and the emitter-field coupling rate $\Omega$ increases. The PF effect on the CRS is greater for the larger number of emitters (compare curves with $n_c=2$ and $N_0=100$ and $200$).

\begin{figure}[thb]\bc
\centering
\includegraphics[width=7.9cm]{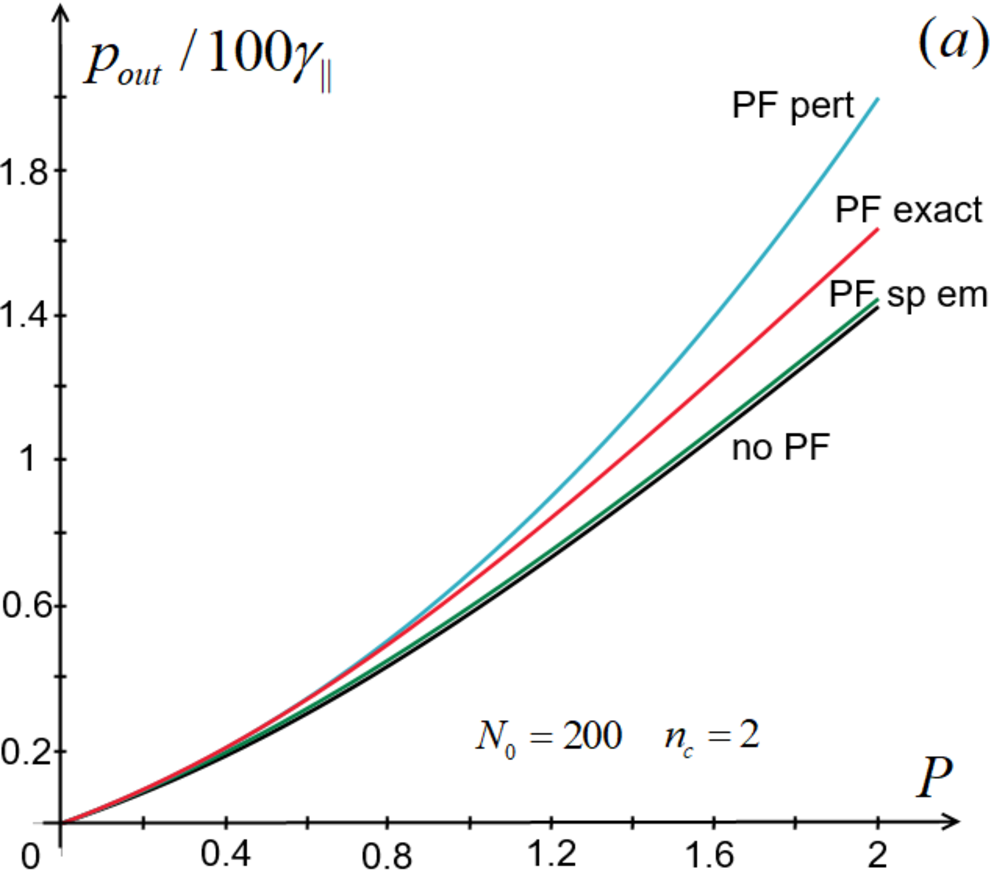}
\includegraphics[width=8.2cm]{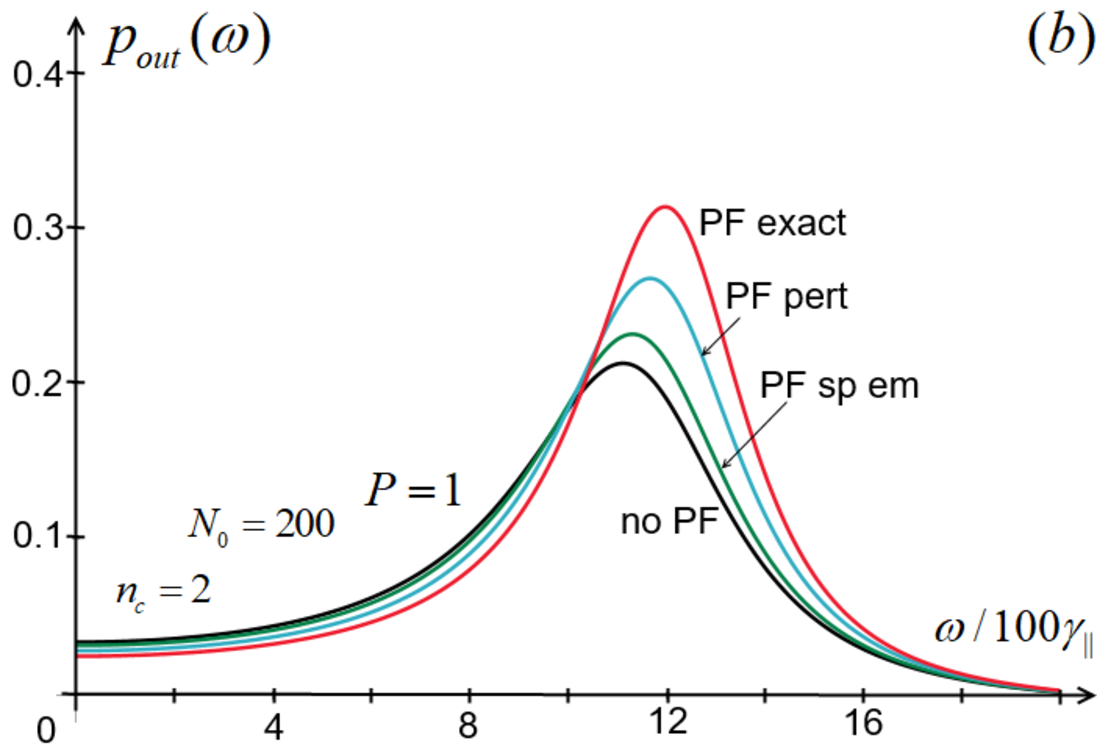}
\caption{The mean (a) and the spectrum (b) of the output power  of non-SR LED found without PF, with PF effect on the spontaneous emission only, and with the perturbative and non-perturbative PF effect on the LED stimulated emission.}
\label{Fig1_1}\ec
\end{figure}

In Figs.~\ref{Fig1_1}, we compare the mean output power $p_{out}(P)$ [Fig.~\ref{Fig1_1}(a)] and the output power spectra $p_{out}(\omega)$ for $P=1$ [Fig.~\ref{Fig1_1}(b)]. These are  found  in various approximations for a non-SR LED with $2\kappa/\gamma_{\perp}=0.05$, $n_c=2$ and $N_0=200$. 
%
%
%
Curves marked by {\em "no PF"} are obtained without PF. Curves marked {\em "PF sp em"} are with the effect of PF on  spontaneous emission only. Curves with {\em "PF pert"} mark are found in the perturbation approach of \ct{PhysRevA.105.053713}. Curves marked by {\em "PF exact"} are found by the non-perturbative approach of this paper. As shown Figs~\ref{Fig1_1}, for non-SR LEDs, we see that   PF has a very small effect on spontaneous emission.  The perturbative approach of Ref. \ct{PhysRevA.105.053713} overestimates the contribution of PF to stimulated emission in Fig.~\ref{Fig1_1}(a), while  the non-perturbative approach of this paper predicts the maximum change in the spectrum $p_{out}(\omega)$  due to PF in Fig.~\ref{Fig1_1}(b).

Now we consider the SR LED with a large adiabatic parameter $2\kappa /{{\gamma }_{\bot }}>1$. Note that Eqs.~\rf{MBE_0} taken without population fluctuations [that is, without  $\delta {{\hat{N}}_{e}}$ in Eq.~\rf{MBE_2}] have a symmetry: the mean output power and the output power spectrum remain the same in the $2\kappa \leftrightarrow {{\gamma }_{\bot }}$ exchange.  Such a symmetry tells us that the coherence stored in the field in the non-SR LED is transferred to the two-level emitter system in the SR LED.  The $2\kappa \leftrightarrow {{\gamma }_{\bot }}$ exchange symmetry is broken when the PFs are taken into account. We take the parameters used in Figs.~\ref{Fig1}, \ref{Fig2} for non-SR LEDs and make the exchange $2\kappa \leftrightarrow {{\gamma }_{\bot }}$. The adiabatic parameter $2\kappa /{{\gamma }_{\bot }}=20$ is large after the exchange, so the LED is superradiant.  

%
%
\begin{figure}[thb]\bc
\centering
\includegraphics[width=7.5cm]{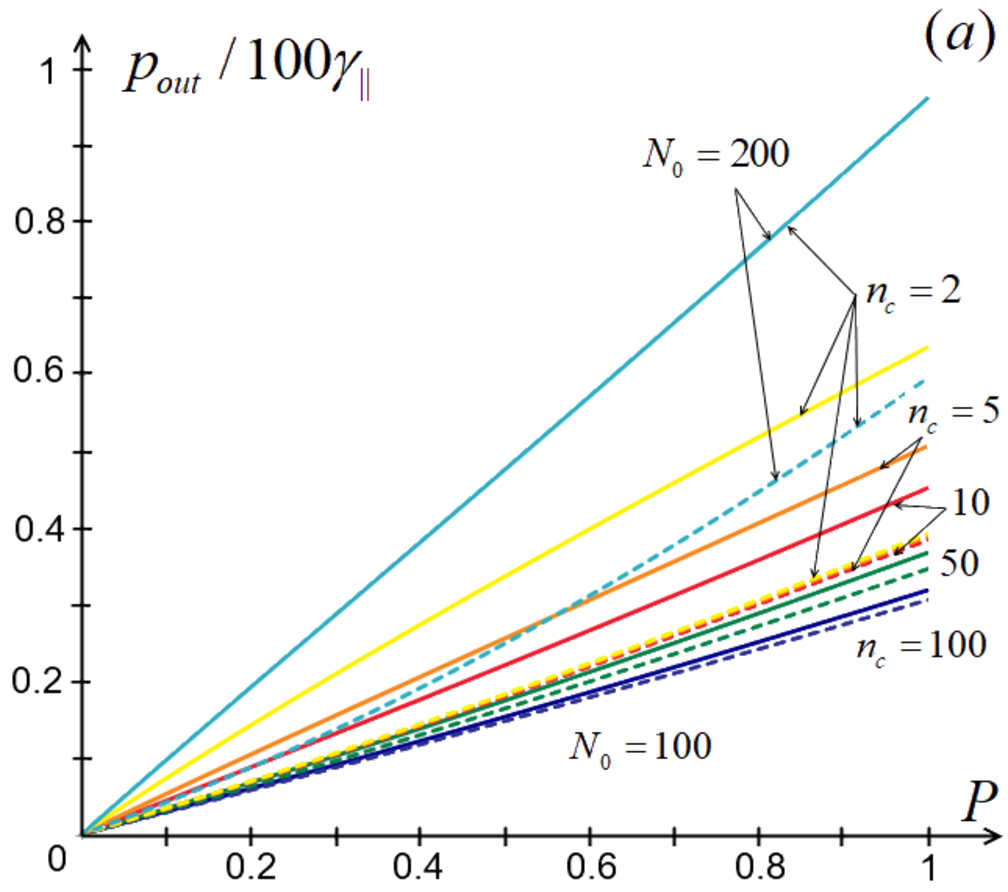}\hspace{1.5cm}\includegraphics[width=7.4cm]{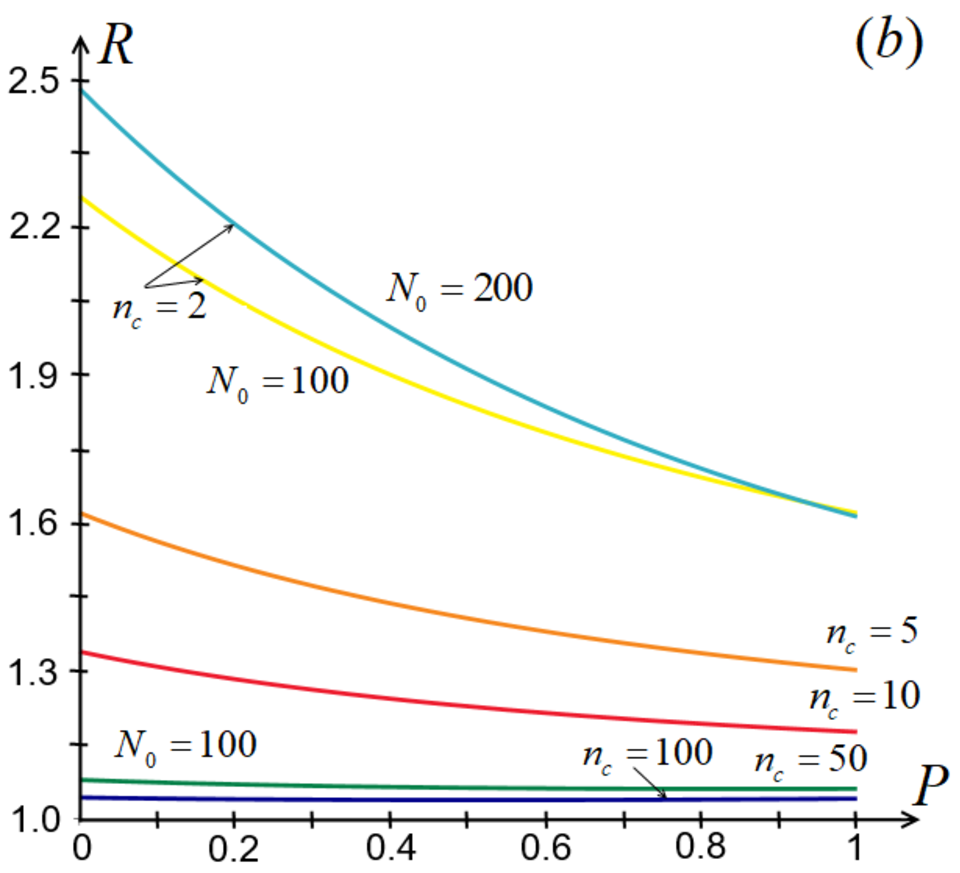}
\caption{(a) The output power ${{p}_{out}}$ (in photons/sec) of the SR LED as a function of the normalized pump $P$.  Parameter values and the curve numbering  are the same as in Fig.\ref{Fig1}(a), but the value of $2\kappa $ is replaced by ${{\gamma }_{\bot }}$ and vice versa. The dashed curves are the same as in Fig.\ref{Fig1}(a) - due to the symmetry to the $2\kappa$  $\leftrightarrow$ ${{\gamma }_{\bot }}$ exchange when Eqs.\rf{MBE_0} are without PF. (b) The increase factor $R$ of the LED output power due to PF.  $R$ and  ${{p}_{out}}$ are larger for SR LED than for non-SR LED on Fig.\ref{Fig1}.  }
\label{Fig3}\ec
\end{figure}
%

Fig.\ref{Fig3}(a) shows the output power $p_{out}(P)$ of the SR LED. Values of parameters are the same as in Fig.\ref{Fig1}(a), except for the exchange $2\kappa \leftrightarrow {{\gamma }_{\bot }}$. The dashed $p_{out}(P)$ curves calculated without PF are the same as in Fig.\ref{Fig1}(a). The solid curves are found with PF, and we see that the output power of SR LED is higher than the power of non-SR LED, when PF are taken into account: compare the solid curves in Fig.\ref{Fig1}(a) and in Fig.\ref{Fig3}(a). 

The relative output power increase  $R$ due to PF for SR LED is shown in Fig.\ref{Fig3}(b) for the same parameters as in Fig.\ref{Fig3}(a). $R$ grows with  $\Omega/\gamma_{\perp}$ and decreases with $P$. 
R for SR LEDs is much higher than for non-SR LEDs, compare the curves with the same $n_c$ in Fig.\ref{Fig1}(b) and Fig.\ref{Fig3}(b). The maximum $R$ is $R=2.5$, see the curve with $n_c=2$ and $N_0=200$.

We observe that $R$ decreases with $P$ in the SR LED [Fig.~\ref{Fig1}(b)],  which differs from the non-SR LED, where $R$  increases with $P$  [Fig.~\ref{Fig3}(b)]. This is because the PF contributes more to spontaneous emission than to stimulated emission in the SR LED. As $P$ increases, the stimulated emission dominates the spontaneous emission, reducing  the relative PF contribution to the emission of the SR LED and, therefore, reducing $R$, as shown in   Fig.~\ref{Fig3}(b). In contrast, PF contributes more to stimulated emission than to spontaneous emission in the non-SR LEDs, even for small $P$. Thus, $R(P)$ increases with $P$, as shown in Fig~\ref{Fig1}(b), when stimulated emission increases with $P$.   

%
%
\begin{figure}[thb]\bc
\centering
\includegraphics[width=9.2cm]{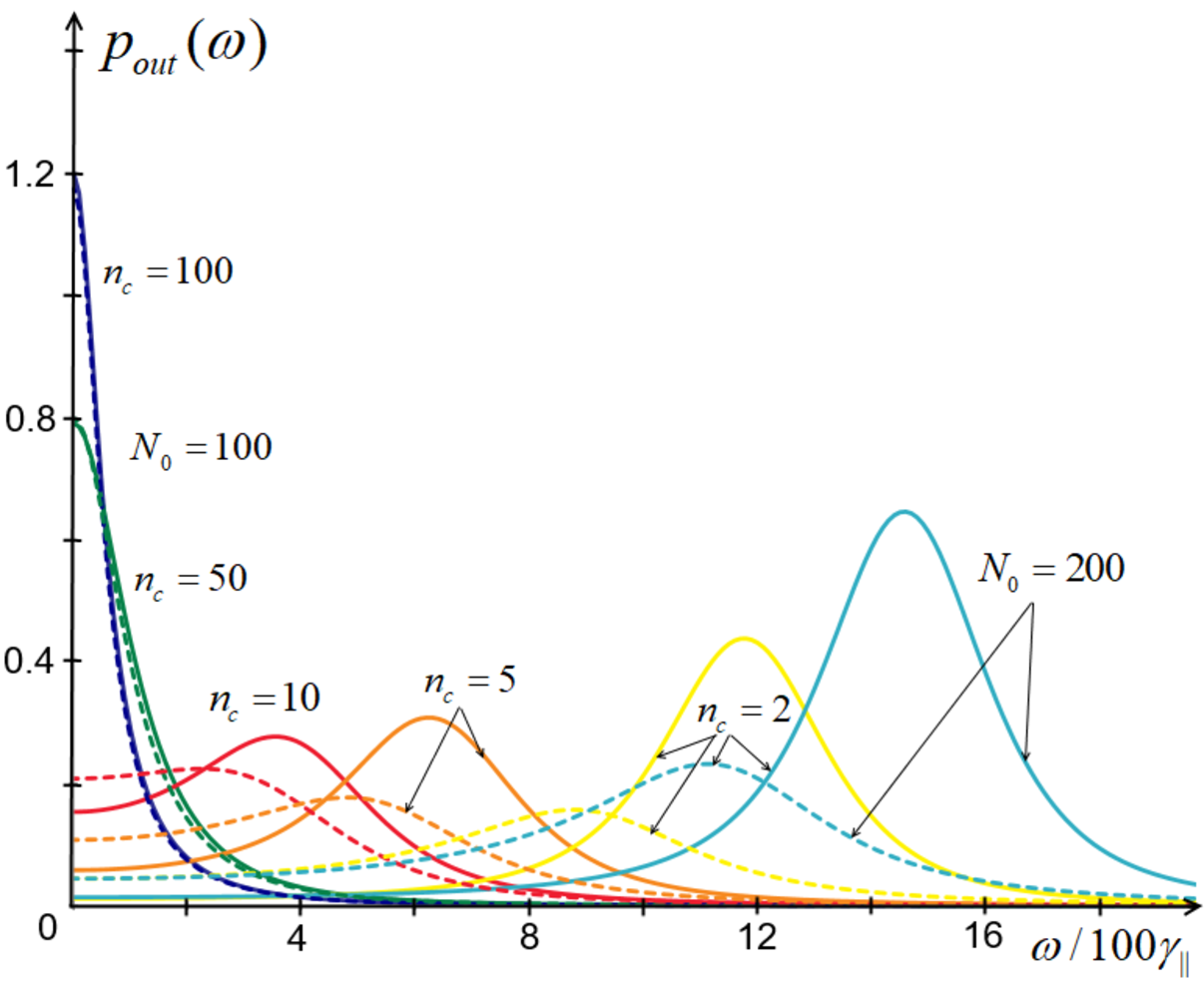}
%
\caption{The output field  spectra for SR LED for $P=1$. Parameters for curves are the same as in Fig.\ref{Fig3}.  Solid (dashed) curves are calculated with (without) PF. Values of $n_c$ are shown near curves. PF increase CRS (shift and enlarge the maxima) much greater than in Fig.~\ref{Fig2} for non-SR LED. The PF effect is greater for larger $N_0$ and smaller $n_c$ (compare curves with $n_c=2$ and $N_0=100,200$).}
\label{Fig2b_rev}\ec
\end{figure}
%
The output power spectra $p_{out}(\omega)$ of the SR LED are shown in Fig.\ref{Fig2b_rev}. Here we see a strong effect of PF on $p_{out}(\omega)$, larger than on the spectra of the non-SR LED in Fig.\ref{Fig2}. The solid curves show that PF  significantly increases the CRS  - relative to the curves found without PF (dashed curves). The increase and the shift in CRS maxima of SR LED spectra is much greater than for non-SR LED: compare Figs.~\ref{Fig2} and \ref{Fig2b_rev}.

\begin{figure}[thb]\bc
\centering
\includegraphics[width=7.5cm]{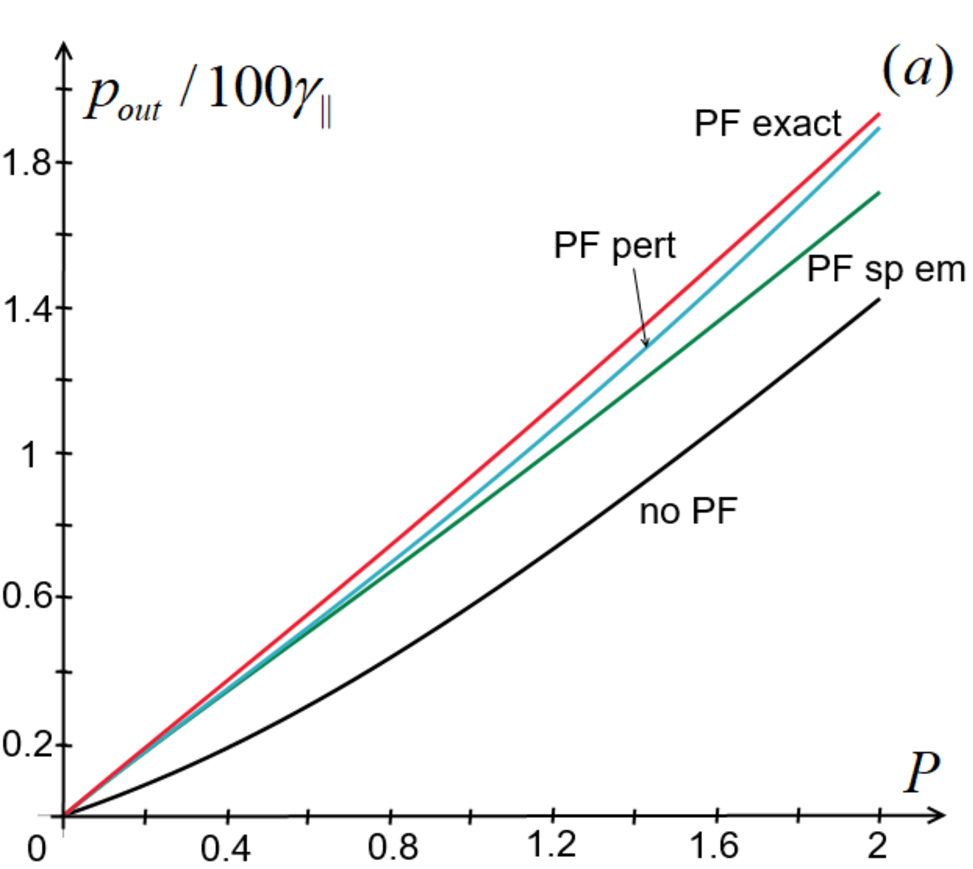}
\includegraphics[width=8.2cm]{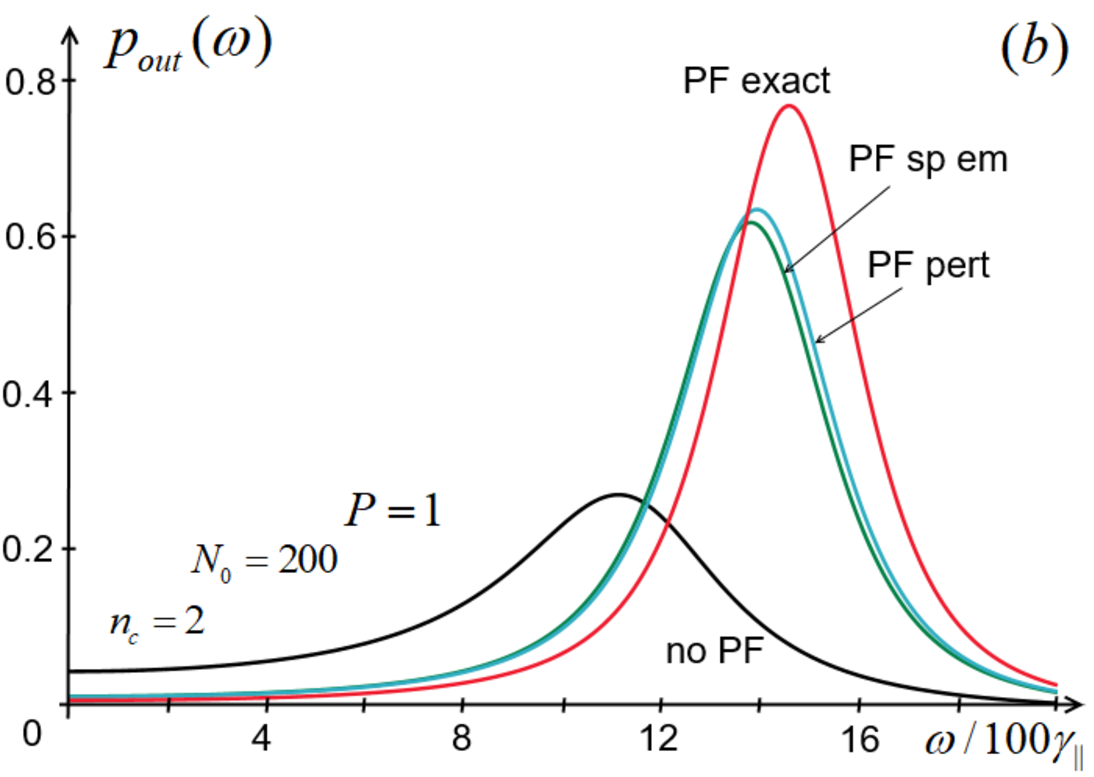}
\caption{The mean (a) and the spectrum (b) of the output power  of non-SR LED found without PF, with PF effect on the spontaneous emission only, and with the perturbative and non-perturbative account of PF effect on the LED stimulated emission.}
\label{Fig1_2}\ec
\end{figure}

The mean output power $p_{out}$ and the power spectrum $p_{out}(\omega)$ found  in Refs. \ct{PhysRevA.105.053713,PhysRevA.105.053713,https://doi.org/10.1002/andp.202400121} and in this paper for SR LED with $n_c=2$  $N_0=200$ and $2\kappa/\gamma_{\perp}=20$ are shown in Figure \ref{Fig1_2}. 
%
%
%
The notations of the curves are the same as in Fig.~\ref{Fig1_1}. We can clearly see the effect of PF on the SR LED spontaneous emission when $P$ is less than 1. This is different from non-SR LED, where the PF effect on the spontaneous emission is small [compare "{\em no PF}" and "{\em PF sp em}" curves in Figs.~\ref{Fig1_1}(a) and \ref{Fig1_2}(a)]. The field spectrum $p_{out}(\omega)$ is significantly affected by PF, much more than it is in non-SR LED, [see Figs.\ref{Fig1_1}(b) and \ref{Fig1_2}(b).] The maximum PF effect on SR spectra is found in the non perturbative approximation of this paper [compare the "{\em no PF}" and "{\em PF exact}" curves in Fig.~\ref{Fig1_2}(b)]. The maximum PF effect on the LED spectra is found in the non-perturbative approach because of a {\em nonlinear} dependence of $n(\omega)$ on $\delta^2N_e$ in Eq.~\rf{n_sp_appr}. This dependence does not appear in the approximations of Refs. \ct{PhysRevA.105.053713,PhysRevA.105.053713,https://doi.org/10.1002/andp.202400121}, which are {\em linear} in $\delta^2N_e$.

\subsection{\label{Sec5} Summary and discussion of results}

We solve quantum nonlinear Maxwell-Bloch equations \rf{MBE_0} below  lasing threshold, when the spectrum of the population fluctuation (PF)  is narrower than other spectra.  The result for the cavity field spectrum $n(\omega)$ is given by Eq.~\rf{n_sp_appr}. This equation is nonlinear in the PF dispersion $\delta^2N_e$, which is different to previous approximations in Refs. \ct{PhysRevA.105.053713, https://doi.org/10.1002/andp.202400121}. Because of the nonlinearity of $n(\omega)$ in $\delta^2N_e$ PF can strongly affect the LED spectra. Using Eq.~\rf{n_sp_appr} we investigate PF as a new source of the LED radiation. We found that PF coupled with the cavity vacuum  (or real) field  contributes to spontaneous (or stimulated)  emission to the cavity mode.  We compare the effect of PF on the radiation of the  non-SR and SR LEDs for parameters, when LEDs have the same output without PF. 
We found that PF increased the output and changed the field spectra  in SR LED much more  than in non-SR LED, see Figs.~\ref{Fig1}, \ref{Fig3} and Figs~\ref{Fig2}, \ref{Fig2b_rev}. We compare the results of different  approximations and see that the strongest effect of PF on the field spectra is predicted by the non perturbative approach of this paper.

PF is more important for SR LEDs than non-SR LEDs when it comes to spontaneous emission [see Figs.~\ref{Fig1_1}(a) and \ref{Fig1_2}(a)]. This is why the relative contribution $R$ of PF is reduced with the pump P in SR LEDs and increased in non-SR LEDs [see Figs.\ref{Fig1}(b) and \ref{Fig3}(b)]. Therefore, the reason for such $R(P)$ is that stimulated emission overcomes spontaneous emission as $P$  increases. 

We see from Eq.\rf{MBE_2} and the polarization Langevin force power spectrum \rf{power_sp_v} that the contribution of the population fluctuations $\delta\hat{N}_e$ to the polarization is proportional to $\Omega$. The  contribution of $\delta\hat{N}_e$ competes with the polarization decay with the rate ${{\gamma }_{\bot }}$. Therefore, the contribution of PF to the radiation increases with the ratio $\Omega /{{\gamma }_{\bot }}$.  This ratio is small for non-SR LEDs, corresponding to the curves in Figs.\ref{Fig1}: $\Omega /{{\gamma }_{\bot }}$ is 0.03, 0.04, 0.1, 0.14 for first four low curves  and 0.22 for two top curves.  Otherwise, $\Omega /{{\gamma }_{\bot }}$  is large for the curves in Figs.\ref{Fig3} corresponding to the SR LED, it is 0.61, 0.87, 1.9, 2.7 for the first four and $\Omega /{{\gamma }_{\bot }}=4.3$ for two top curves. 
Thus, a large medium-field coupling
$\Omega$, on the order of ${{\gamma }_{\bot }}$ or larger, and relatively small $\gamma_{\perp}$ are required for a significant contribution of the PF to the LED radiation.  If we consider a non-SR LED with $2\kappa \ll {{\gamma }_{\bot }}$, exchanging $2\kappa \leftrightarrow {{\gamma }_{\bot }}$ thus coming to an SR LED after the exchange, we see no difference in the output of such LEDs without considering PF.

Taking PF into account, we see that the PF contributes more to the emission of the SR LED (with a small $\gamma_{\perp}$) than to the emission of the non-SR LED (with a large $\gamma_{\perp}$). The contribution of PF to spontaneous emission depends on the ratio $\sim \Omega^2/\gamma_{\perp}\gamma_{\parallel}$ of the second (PF) and the first (polarization) terms in the power spectrum \rf{power_sp_v}. This ratio increases with smaller $\gamma_{\perp}$. This is why PF's contribution to spontaneous emission is greater for SR LEDs with smaller $\gamma_{\perp}$ than for non-SR LEDs with larger $\gamma_{\perp}$.

The absolute value of PF contribution depends on $\Omega$ and the number of emitters ${{N}_{0}}$. The value of ${{N}_{0}}$ does not change much the increase factor $R$; see two top curves in Figs.\ref{Fig1}(b) and \ref{Fig3}(b). For values of $\Omega$ approaching the maximum (i.e., for the LED cavity size   $\sim\lambda /2$), the increase in LED output power due to PF is approximately $R=2.5$, relative to the radiation power found without PF; see two top curves  in Fig.\ref{Fig3}(b). In Figs.~\ref{Fig2}, we can see that PF increases the CRS more for SR LED than for non-SR LED. Thus, PF  significantly increases the efficiency of SR LEDs and CRS, when $\Omega /{{\gamma }_{\bot }}\sim 1$ or greater. 

The non-SR LED and SR LED outputs are the same without PF if all LED parameters are the same except for the $2\kappa\leftrightarrow \gamma_{\perp}$ exchange. However, with these parameters, the SR LED output is much larger than the non-SR LED output with PF; compare Figs.~\ref{Fig1}(a) and \ref{Fig3}(a). This is because of the  $\Omega/\gamma_{\perp}$ ratio is greater for the SR LED than for the non-SR LED,  providing a greater contribution ($\sim \Omega/\gamma_{\perp}$) of PF to the  SR LED radiation, as explained above.

Experimental investigations, for example, Refs. \ct{5420237,PhysRevLett.52.341} of population fluctuations (pump noise \ct{5420237} or multiplicative noise \ct{,PhysRevLett.52.341}) have focused on the photon statistics rather than the PF effect on the LED power. Large field intensity fluctuations observed in SR laser experiments \ct{Jahnke} can, in principle, be explained by the PF effect \ct{https://doi.org/10.1002/andp.202400121}. As far as we know, no experiments have been performed to investigate the PF effect on the power of small LEDs. We suggest conducting such an experiment. This can be achieved using photonic crystal LEDs with strong medium-field coupling in a small cavity of the size about the wavelength. For example, one could use two LEDs, as in examples above, with $\kappa = 2.5\cdot 10^{10}$~rad/s (cavity quality factor $Q=2.4\cdot 10^4$) and $5\cdot 10^{11}$~rad/s ($Q=1.2\cdot 10^3$). The high-Q cavity LED operates at room temperature. Therefore, its $\gamma_{\perp}$ is approximately $10^{12}$~rad/s and $2\kappa/\gamma_{\perp}$ is much less than $1$. It is, therefore, a non-SR LED. The second low-Q cavity LED is cooled  to a temperature of about $T=10$ - $15$~K. Assuming $\gamma_{\perp}\sim T$, it has $\gamma_{\perp} \sim 5\cdot 10^{10}$~rad/s. Thus an exchange of $2\kappa$ and $\gamma_{\perp}$ occurs between the two LEDs, making the low-Q cavity LED superradiant. Our  zero-order approximation  indicates that outputs of the two LED's should be similar if PF does not affect LEDs. An experimental observation of an increase in the output of the low-Q cavity LED, relative to the high-Q cavity LED will demonstrate the PF effect in the LED output.  

In the future we will study $n(\omega)$, which is nonlinear in $\delta^2N_e$ according to Eq.~\rf{n_sp_appr}. We will also study the mean photon number $n$ for a strong pump $P>1$, when $\delta^2N_e$ may depend on $n$. This kind of  nonlinearity $n(\delta^2N_e(n))$ may lead to  interesting results, such as  resonances, in the dynamics of LEDs and lasers.

Another interesting task for the future is to model the LED regime with a large $P>1$, when the widths of all spectra are of the same order $\kappa \sim \gamma_{\perp} \sim \gamma_{\parallel}$. In this case, the approximation of Eq.~\rf{lf_spectrum} by Eq.~\rf{n_sp_appr} is invalid. So we have to develop a numerical procedure for finding $n(\omega)$ from the integral equation \rf{n_sp_appr}. The population fluctuation power spectrum $\delta^2N_e(\omega)$ in Eq.~\rf{n_sp_appr} can be found using the quantum regression theorem and the Fourier-component operator $\delta\hat N_e(\omega)$ determined from Eq.~\rf{MBE_3}. 

\section{\label{Sec6} Conclusion}

We solve the quantum nonlinear Maxwell-Bloch laser equations  in the LED regime, when the population fluctuation power spectrum is much narrower than the field and the polarization spectra. We consider population fluctuations (PF) in the  equations using a non perturbative approach and show that PF significantly increase the LED output power: up to 2.5 times under certain conditions. 
The output field spectra of the SR LEDs are affected by PF.
Specifically, PF increases the collective Rabi splitting
\ct{Andre:19}. The maximum power increase due to PF occurs for the super radiant LEDs when the field-medium coupling (the Rabi frequency) is of the order of or greater than the  polarization decay rate. The approach developed in this paper can be used to study of miniature lasers, plasmonic devices \ct{PROTSENKO2024101297}, and nonlinear quantum optical devices \ct{PhysRevA.108.023724}.       
 
\medskip

\bibliography{myrefs}

\end{document}